\documentclass[12pt]{article}

\usepackage[semicolon, authoryear]{natbib} 
\usepackage{graphicx}
\usepackage{times}
    
\usepackage{url}  
\usepackage{hyperref}

\usepackage{makeidx}
\usepackage{graphicx}
\usepackage{amsmath} 
\usepackage{verbatim}

\begin{document}

	\def\spacingset#1{\renewcommand{\baselinestretch}%
		{#1}\small\normalsize} \spacingset{1}

	\title{ Homeostatic behavioural response to COVID-19 infections returns R to a set-point of 1\\}
	\author{ \\ Fintan Costello \\ School of Computer Science,\\ University College Dublin\\ \\
		Paul Watts \\ Department of Theoretical Physics,\\National University of
		Ireland  Maynooth\\ \\ 
	Rita Howe \\ School of Public Health, Physiotherapy and Sports Science,\\ University College Dublin\\}

	\thispagestyle{empty}
	\vspace{2cm}

	\bigskip

	\def\spacingset#1{\renewcommand{\baselinestretch}%
		{#1}\small\normalsize} \spacingset{1}

\maketitle 
 
\pagebreak
  
\begin{abstract}
One clear aspect of behaviour in the COVID-19 pandemic has been people's focus on, and response to, reported or observed infection  numbers in their community.	We describe a simple model of infectious disease spread in a pandemic situation where people's behaviour is influenced  by the current risk of infection and where this behavioural response acts homeostatically to return infection risk to a certain preferred level.  This model predicts that the  reproduction rate $R$ will be centered around a median value of 1, and that a related measure of relative change in the number of new infections  will follow the standard Cauchy distribution.  Analysis of worldwide COVID-19 data shows that the estimated reproduction rate has a median of 1, and that this measure of relative change calculated from reported numbers of new infections closely follows the standard Cauchy distribution at both an overall and an individual country level. 
\end{abstract}

 In epidemiological models of disease spread, infection numbers at time $t$ are a function of disease transmissibility,  incubation and recovery rates (all fixed properties of the disease), of the proportion of infectious and susceptible individuals in the population at time $t$ (functions of the state at time $t-1$), and of behaviour: in particular, of the average number of contacts individuals make with others at that time, $K_t$.  In some models \citep{bertozzi2020challenges} this contact number $K_t$ is taken as to be constant, giving a fixed transmission rate of $\beta$; in others $K_t$ (or $\beta_t$) is  treated as a free parameter, varying with time in a way that is not described within the epidemiological model but instead is estimated via fitting the model to data \citep{ndairou2020mathematical,ihme2020modeling} or by using mobility or contact tracing datasets \citep{nouvellet2021reduction,badr2020association,russo2020tracing}.   
 
  We give a simple  model of  how people's behaviours (and so contact numbers) change over time in response to their assessment of risk of infection at that time.   
In this model people's behavioural response to infection balances the risk of infection associated with contact against the various  (economic, social and psychological) gains associated with contact.  We assume that people can estimate their risk of infection given a certain number of contacts (a risk that depends on infection rates in the community) and that each person has a certain constant risk or probability of infection per day, $x$, which they are willing to accept (whose value depends on their age, health, financial status, and so on).  Each person will set their number of contacts on a given day so that, based on their estimate of the risk per contact, their overall risk that day is approximately $x$ (so maximising their gains from contact without incurring unacceptable risk).    We assume that actors such as businesses or governments will behave in a similar way, balancing risk against gain in setting policy responses to infection.   In this model people will tend to change their behaviour so that their probability of infection varies around $x$, reducing contacts when risk is higher than $x$ but increasing contacts when it is below $x$.   Since the population risk of infection is the average of all individual risks, the overall probability of infection will vary over time around some constant $X/N$  (where $X$ the sum of acceptable risk levels and $N$ the population size), and so the expected number of new infections per day will vary around $X$.  Finally, since the reproduction rate $R$ is the number of new infections caused by an existing infected individual, with new infections varying around a constant this model predicts that $R$ will vary around a median value of $1$.

Two aspects of this model may be surprising.  First, it goes against the common understanding that `an R above one means an outbreak is growing, and below one means that it is shrinking' \citep{adam2020guide}.   In this model an $R$ below $1$ does not necessarily mean the outbreak is shrinking: instead an $R$ below $1$ leads to an increase in contact numbers, which can cause a subsequent increase in new infections and in $R$.   Second, this model assumes that people are able to accurately judge the probability of infection and adjust their behaviour appropriately, contradicting the common view that  `In making predictions and judgments under uncertainty, people do not appear to follow the calculus of chance or the statistical theory of prediction.  Instead they rely on a limited number of heuristics which sometimes yield reasonable judgments and sometimes lead to severe and systematic errors' \citep{TverskyKahneman1973}.    This aspect of the model is motivated by our previous work suggesting that people's assessment of probability do in fact follow the statistical theory of prediction, and that observed patterns of systematic error in judgement are caused by the regressive effects of random variation or noise \citep{costelloWatts2014,CostelloWattsConditionals,costello2018invariants,costello2019illusory,HoweCostello2020}.  

  \begin{figure*}[t!h!] 
	\begin{center}
		\scalebox{0.8}{\includegraphics*[viewport= 0 0  550 350]{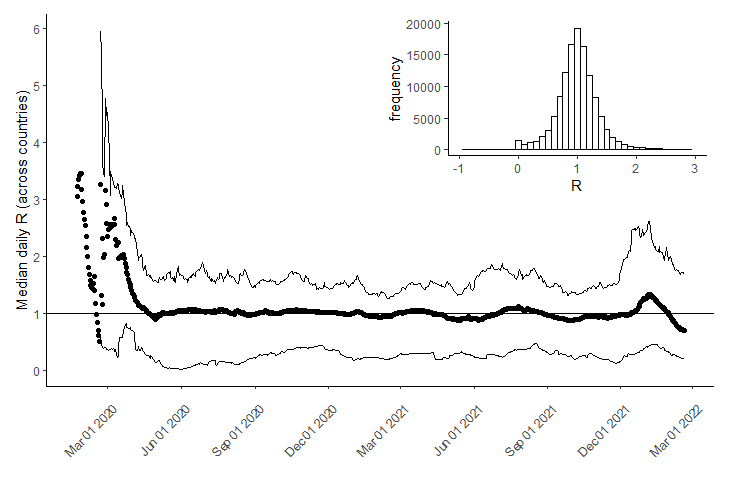}}
	\end{center} 
	\caption{ Median of $R$ values on each day (points) with $2.5\%$ and $97.5\%$ percentiles for country $R$ values on each day (lines). The inset shows a histogram of $R$ values (bin size $0.1$).   There are no percentile lines before $21$ February $2020$, because only $R$ values for China are reported before that date. }
\end{figure*}

We tested this model using data from the Our World in Data COVID hub  \citep{owidcoronavirus}  (accessed February $22$, $2022$).  This dataset gives the number of new COVID-19 infections reported each day for $225$ countries, from the Johns Hopkins University COVID-19 Data Repository  \citep{dong2020interactive};  the  reproduction rate each day for $187$ countries, estimated using a Kalman Filter approach \citep{arroyo2021tracking}; and the estimated stringency of government pandemic response each day for $173$ countries, from the Oxford COVID-19 Government Response Tracker \citep{hale2021global}.   The median estimated $R$ across this dataset was $1.0$ (Fig $1$) with median estimated $R$s for each individual country being  indistinguishable from $1$ in a one-sample t-test ($t(186)=1.77,p = 0.08$, $95\%$ confidence interval for the mean: $0.95 \ldots 1.003$).\footnote{Analysis code at \url{https://osf.io/7jx64/?view_only=a092078d4745484aab4990a6f6fe9618}.\\ }   

 The reproduction rate  $R$ on a given day $t$ is a function of $n_t$, the reported number of new cases on that day. Could this $R \approx 1$ result be an artefact of the $n_t$ reporting process?  One problem with COVID case numbers is the frequent reporting of $0$ new cases: just under $25\%$ of $n_t$ values in the dataset were $0$, with these often indicating that no reporting took place that day: a number of countries had reliable patterns of $n_t=0$ on weekend days only.  These reporting gaps are visible as a spike in $R$ values at $0$ in the Fig $1$ histogram of $R$ values.   To eliminate these reporting gaps we reran our analysis on a cleaned dataset including only days with $n_t>0$. The median estimated $R$ for $n_t>0$ was $1.02$, with median estimated $R$s for each individual country being  indistinguishable from $1$ in a one-sample t-test ($t(186)=1.1, p = 027$, $95\%$ confidence interval: $0.99 \ldots 1.03$).  All subsequent analyses use this cleaned dataset.
 
Perhaps this result could be caused by a relationship between $R$ and the number of tests being carried out?  If testing increases when $R$ is high, tests would include more cases likely to be negative and would reduce the apparent value of $R$; similarly, low test numbers could increase the apparent value of $R$.  We calculated the correlation between $R$ and number of daily tests carried out in the cleaned dataset. There was no significant difference between median $R$ values for countries where this correlation was positive and those  where it was negative ($t(126.96) = 0.36,p- = 0.72$).

Perhaps this result is a consequence of government interventions alone, rather than behavioural responses to risk?  To check this we compared median $R$ values for countries where the average government stringency level was above the overall mean stringency, and those for which it was below.  There was no significant difference between median $R$ values for these groups ($t( 157.43) = 1.74, p= 0.08$); the average median $R$ was slightly higher in the high-stringency group ($1.02$) than the low-stringency group ($1.0$).

These results suggest that the number of new infections at time $t$, $i_t$, varies around some constant $X$ as a consequence of people's behavioural response to  infection risk (and so $R$ is distributed around $1$).  What can we say about the distribution of these values $i_t$?  This behavioural response has a natural lag, $L$, which represents the time between an infection occurring (at time $t-L$, say) and that infection being observed by others and causing a behavioural response  (at time $t$). This lag falls somewhere between the incubation and recovery period for the infection (an infection becoming observable only after incubation, and not being observable after recovery), and means that the observed rate of new infections  at time $t$ is equal to the actual rate of new infections at time $t-L$.  If $i_{t-L} > X$ then the overall behavioural response  at time $t$ will  reduce contact numbers, pushing $i_t$ downwards, while if $i_{t-L} < X$ then the overall behavioural response  at time $t$ will increase contact numbers, pushing $i_t$  upwards, and so the difference $i_t - i_{t-L}$  varies around $0$.  Since this overall behavioural response is the sum of all individual responses in the population, from the Central Limit theorem this difference $i_t - i_{t-L}$ will follow a Normal distribution  $ i_t - i_{t-L}  \sim \mathcal{N}( 0, \sigma_t^2)$ with some variance  $\sigma^2_t$ (which may change over time).   The difference $i_{t-2L} - i_{t-L}$ will follow the same distribution (albeit with variance $\sigma_{t-L}^2$). 

 Defining a measure of relative change  in new infection numbers from time $t-2L$ to time $t$,
 \[ D_L(t)= \frac{(i_{t} - i_{t+L}) - (i_{t-2L} - i_{t-L})}{(i_{t} - i_{t-L}) + (i_{t-2L} - i_{t-L})} = \frac{i_{t} - i_{t-2L}}{i_{t} + i_{t-2L} - 2i_{t-L}}   \]
 we see that $D_L$ is the ratio of  two standard Normal variables (sums of common standard deviations cancelling) and so this measure $D_L$ will follow the standard Cauchy distribution $C$ (with location $0$ and scale $1$) for $L$ between the incubation and recovery times.      Assuming that changes in $n_t$ are proportional to changes in $i_t$, we predict that $D_L$ values calculated from reported number of positive tests $n_t$ will also follow the standard Cauchy distribution $C$.

   To test this prediction we compare $D_L$ values calculated for our cleaned dataset  against the theoretical distribution $C$.  For each country, at each day $t$ we calculated $D_L(t)$ for various values of $L$.  For some days $D_L(t)$ could not be calculated (because one of the component infection numbers was missing), or involved division by $0$; these values of $D_L(t)$ were dropped from analysis.    Figure $2$ (inset) shows a probability-probability plot comparing the cumulative probability of $D_L$  for $L=6$ against that of $C$.  Correlation of cumulative probabilities is a measure of goodness of fit between observed and theoretical values; here the correlation was high ($r=0.9997$).     Since probability-probability plots overweight extreme values, we also analysed the relationship between $C$ and  $D_L$  for values near the midpoint of the range, by selecting the subset of $D_L$ values between $-15$ and $15$ (over $95\%$ of the total sample). Figure $2$ (main) shows a histogram of these values.   The correlation between $D_L$ and $C$ values for this central-region histogram was  $r=0.993$.    As an additional check we calculated location and scale estimates by taking the median of  $D_L$  and the median of the absolute value of $D_L$; these values were $0.02$ and $1.02$ respectively, confirming the fit to the standard Cauchy distribution.
   
     \begin{figure*}[t!h!] 
   	\begin{center}
   		\scalebox{0.8}{\includegraphics*[viewport= 0 0  550 350]{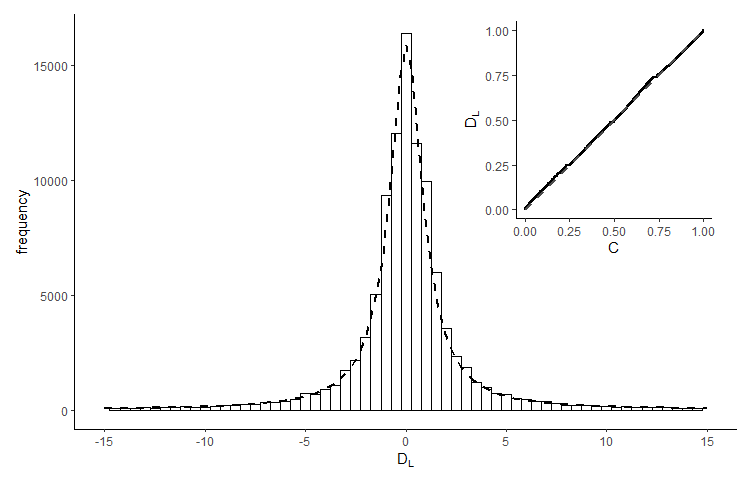}}
   	\end{center} 
   	\caption{ Histogram of $D_6$ calculated from the cleaned dataset in the central $-15\ldots 15$ range (bin size $0.5$) with standard Cauchy distribution $C$  (dashed line, $C$ distribution scaled by bin size and total histogram frequency for comparison).   The inset shows a probability-probability plot comparing  theoretical and observed cumulative probabilities across the entire range: the solid line in that plot is actually made up of $88,832$ points, one for each $D_L$ value calculated in the dataset: the dashed diagonal line (mostly hidden by these points) is the line of identity between theoretical and observed cumulative probabilities. }
   \end{figure*}
   
 We  carried out the same analysis for individual country data for $L=6$.  Correlations between cumulative probabilities of $C$ and  $D_L$  for individual countries were greater than $r=0.98$ for all  countries; for binned $D_L$ and $C$ values in the $-15 \ldots 15$ region the mean correlation was $r=0.97$ (with a $2.5\% \ldots 97.5\%$ percentile range of $0.83$ to $0.99$). The $2.5\% \ldots 97.5\%$ percentile range for $D_L$ medians (location parameter estimates) across countries was $-0.06 \ldots 0.22$ with an overall median of $0.03$, for absolute medians (scale parameter estimates) the same range was $0.71 \ldots 1.38$ with an overall median  of $1$, agreeing with the predicted values of $0$ and $1$ for location and scale.  The standard Cauchy distribution gives a good fit to values of  $D_L$ for individual country data.

We predicted that this agreement should hold for $L$ approximately between $5$ and $14$ (estimates for the incubation and recovery period for COVID-19).  To test this we carried out the above analysis for all values of $L$ from $2$ to $100$.  Probability-probability correlations, slopes and intercepts did not change noticeably because changes in $L$ primarily affect $D_L$ around the median.  Figures $3$ plots the correlation  between binned values of $D_L$ and $C$ values  in the $-15\ldots 15$ median region, for each value of $L$.  Correlation was highest for values of $L$ in the $5$ to $14$ region, as predicted, supporting the behavioural response model.   
 
   \begin{figure*}[t!h!] 
 	\begin{center}
 		\scalebox{0.8}{\includegraphics*[viewport= 0 0  480 350]{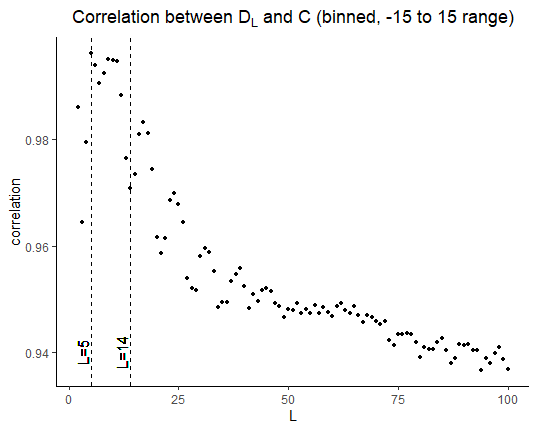}}
 	\end{center} 
 	\caption{ Each point shows the correlation $r$ between values of $D_L$ in the $-15\ldots 15$ central range and the standard Cauchy distribution density for those bins,  for values of $L$ in from $2$ to $100$.  }
 \end{figure*} 

 There are a number of clear limitations to these results.  First, our model of homeostasis due to behavioural response assumes that the susceptible population is aware of and responding to the risk of infection, and so applies to epidemic or pandemic situations only: we do not expect this homeostatic effect to hold in narrower outbreak situations.  Second, our model depends on the assumption that new infection numbers at time $t$ are a reflection of the probability of infection at that time.  This assumption holds for infections with short incubation and recovery periods; for infections where these periods are longer, this assumption doesn't hold. Third: our model assumes that people are free to limit their number of contacts to match their acceptable level of risk.  For some demographics this is not the case: people in poverty, for example, may be economically unable to limit their contacts in this way, and so will have an estimated risk of infection systematically above their acceptable risk level.  Assuming that people's acceptable risk levels are well-calibrated, this predicts increased infections in such demographics \citep{patel2020poverty,little2021impact}.  Fourth: we assume that reported infection numbers are proportional to actual infection rates.  If reported infection numbers do not follow actual infection numbers, we do not expect these results to hold.   
  
 A final caveat concerning the interpretation of these results.   At first glance our results may suggest that government responses to infection have no value or no effect.  This is not the case: government restrictions on contact clearly act to reduce the risk of infection.   Instead, our model suggests that  government restrictions  act to reduce the overall level of acceptable risk $X$, so reducing the number of infections to a lower (but still approximately constant) value.  Relaxation of those restrictions then produces an increase in the overall level of acceptable risk, $X$, and so causes infection rates to rise, but only to that new level.

 \bibliographystyle{apalike}
 \bibliography{references}

\end{document}